\documentclass[journal=jpccck,manuscript=article,linenumber]{achemso}
\usepackage[utf8]{inputenc} 
\usepackage[version=3]{mhchem} 
\usepackage{marginnote}
\usepackage{color}
\usepackage{soul}
\usepackage{graphics}
\usepackage{graphicx}
\usepackage{amsmath}
\usepackage{amssymb}
\usepackage{bm}
\usepackage{dcolumn}

\usepackage{epstopdf}

\author{Raul Esquivel-Sirvent}
\email{raul@fisica.unam.mx}
\affiliation
{Instituto de F\'{i}sica, Universidad Nacional Aut\'onoma de M\'exico, Apdo. Postal 20-364, M\'exico D.F. 01000, M\'exico}

\title{Anomaly of the dielectric function of water  under confinement and its role in van der Waals interactions}

\begin{document}

\begin{abstract}
We present a theoretical calculation of the changes in the Hamaker constant due to the anomalous reduction of the static dielectric function of water.  Under confinement, the dielectric function of water decreases from a bulk value of 79 down to 2. If the confining walls are made of a dielectric material, the Hamaker constant reduces almost by 90\%.  However,  if the confinement is realized with metallic plates, there is little change in the Hamaker constant. Additionally, we show that confinement can be used to decreases the Debye screening length without changing the salt concentration. This, in turn, is used to change the  Hamaker constant in the presence of electrolytes. 
\end{abstract}


\section{Introduction}
Confinement of liquids in sub micron structures changes their chemical and physical properties \cite{Alcoutlabi_2005}, such as  density,  surface tension \cite{Takei}, as well as  melting and freezing temperatures \cite{Knight}.   The interest in confinement is because of its frequent occurrence in biological systems, geological samples like nano pores in rocks \cite{WANG20141}.  It is the influence of the confining walls that is responsible for the modification of the physical and chemical properties \cite{thompson}. 

An important property of liquids also  affected by confinement, is their dielectric function ($\epsilon$)  \cite{PhysRevLett.109.107801,PhysRevLett.111.089801,PhysRevE.97.012131}. In the case of water, confinement has two effects. First, the static dielectric function decreases, and secondly the dielectric function of water becomes anisotropic.  Several theoretical works have addressed this problem. Using atomistic molecular dynamics simulations (MDS) for water confined between parallel walls, this  decrease of the dielectric properties has been calculated \cite{Schlaich} and the limited movement of the water dipoles is responsible for the change in dielectric function.  Electrostatic interactions between the water molecules and the confining walls do not play an essential role in the changes of $\epsilon$, as shown using MDS calculations of water in closed nanocavities  \cite{Senapati}. 
The need for understanding the reduction of $\epsilon$ has been a topic of intense debate. The role of the orientation of the hydrogen bonds on the dielectric function is determined using MDS to show the reduced rotational diffusion of the water molecules next to the surface \cite{Varghese}.
Experimentally, the change in the dielectric function has been reported by 
Fumagalli \cite{Fumagalli}.  Measurements of water confined between two atomically flat surfaces, show that the perpendicular dielectric function goes from its bulk value $\epsilon_{\bot}=80$ down to an anomalous value of $\epsilon_{\bot}=2$ when the confining walls are at a distance $L=1$ nm.  Contrary to what is predicted, the bulk value is recovered when $L\sim600$ $nm$.  The reduction is consistent with the mobility of the water dipoles near the surfaces and of the hydrogen bond contributions.  

The changes in the dielectric function have significant consequences on the Van der Waals interaction at the nanoscale \cite{Rudy,RevModPhys.88.045003}, in wetting \cite{HOUGH19803} and self-assembly \cite{Young2240}. The theory of generalized  VdW forces developed by Lifshitz \cite{Lif:56}shows that both the spectral dielectric data of the interacting bodies and the medium between them is needed to find the Hamaker constants. Modern density functional theory can also be used to compute the Hamaker constants  \cite{dft}, but using the continuum model to describe the dielectric function of water and of the confining walls is enough to use Lifhitz theory in many situations \cite{Raul1,Raul2} of interest in problems of self-assembly, for example. 
Electrolytes also change the van der Waals interactions \cite{Ningham72,NINGHAM97,Misra,C0SM00040J,Quesada,Jancovici,Petrache7982} and in particular affects the zero frequency Matsubara contribution due to an electrostatic screening that depends on the Debye wavelength. Changes in the value of the dielectric function of the solvent is expected to also have a strong influence on the VdW force.  

In this paper, we present a theoretical calculation of the effects that the anomalous behavior of water has on the Hamaker constant between two flat surfaces in water including the changes induced by the presence of electrolytes. This calculation is based on the Lifshitz theory of dispersive interactions and the experimental results of the confined dielectric function of water reported by Fumagalli \cite{Fumagalli}.

We consider two parallel flat surfaces separated a distance $L$.  Both plates are made of the same material and the interspace between them is water as shown in Fig. (1 a).  As is common  practice in Lifshitz theory, we will work in the complex frequency space. What this entails is the following,  the dielectric function is evaluated making an analytic continuation of the frequency $\omega$ to the complex plane $\omega \rightarrow i \zeta_n$ where the Matsubara frequencies are defined as $\zeta_n=2 \pi KT n/\hbar$ with $n=0,1...$,  the temperature is  $T$, and $K$ is the Boltzman constant. Thus, we make the change $\epsilon(\omega)\rightarrow \epsilon(i\zeta_n)$.  At room temperature $\zeta_n= \times 3.94\times10^{13}$ $rad/s$.

\begin{figure}[h]
\centering
  \includegraphics[height=7cm]{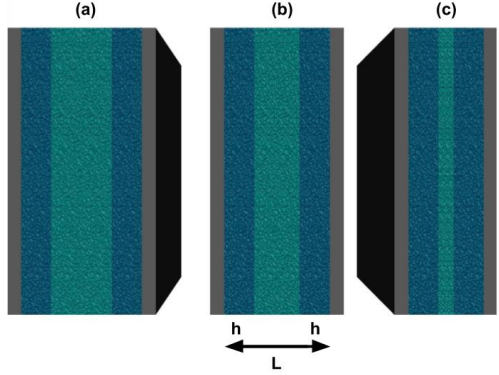}
  \caption{(a) Two parallel plates separated by a distance $L$. Close to the surfaces, the water dipoles loose mobility in a layer of thickness $h$. As the confining plates moves closer together (b) and (c), the perpendicular static dielectric function reduces to a value $\epsilon_{\bot}=2$, as reported in recent experiments \cite{Fumagalli}. }
  \label{static}
\end{figure}

The dielectric function of the plates is $\epsilon_s(i\zeta_n)$ and that of water $\epsilon(i\zeta_n)$. For water in bulk the dielectric function is given by a sum of Lorentz oscillators \cite{PARSEGIAN1981285,FERNANDEZVAREA2000394}
\begin{equation}
\epsilon(i\zeta_n)=1+\frac{d}{1+\zeta_n \tau}+\sum_{i=1}^l\frac{f_i}{\omega_i^2+\zeta_n^2+\zeta_n g_i},
\label{water}
\end{equation} 
where the parameters $\omega_i$, $d$, $g_i$ and $f_i$ were taken from Parseguian \cite{PARSEGIAN1981285}. 

As shown by the experimental work if Fumagali, the confinement induces an anisotropy in the static dielectric function (zero frequency) of water corresponding to a uniaxial material of dielectric tensor 
\begin{equation}
\tilde \epsilon(0)=\begin{pmatrix} 
     \epsilon_{||} &  0 &0 \\
    0  &   \epsilon_{||} &0\\ 
    0& 0&\epsilon_{\bot}   
\end{pmatrix}.
\end{equation}

For higher frequencies ($n\geq1$) the dielectric function is given by Eq. (\ref{water}), and for the next Matsubara frequency, ($n=1$) the value drops to $\epsilon(\zeta_1)=2.08$.  For reference, at room temperature the first Matsubara frequency is $\zeta_1=3.94\times10^{13}$ rad/s $=0.02$ eV, close to the visible range. The basic assumption is that confinement only affects the static dielectric function.

Molecular dynamics simulations suggest that the freezing of the water molecules happens in a layer of a few {\AA}ngstroms close to the surface, while experiments shows that the anomalous behavior begins for separations of $L\sim600$ nm.  As shown is Figure (1) the water between the walls is divided in three sections, the two regions close to the confining surfaces of thickness $h$ and a region of thickness $L-2h$ where the dielectric function of water changes. As the confining separation decreases, the layers where there is a strong change in the dielectric function have a larger statistical weight (Fig. 1(b,c)),  and  the capacitive measurements of $\epsilon_{\bot}(0)$ can be  described by an effective medium  given by a harmonic average \cite{Fumagalli}
\begin{equation}
\epsilon_{\bot}(0)=\frac{L}{2h/\epsilon_i+(L-2h)/\epsilon_{bulk}},
\label{effectivebot}
\end{equation}
where $h=9$ {\AA}ngstrom is the thickness of the near surface layer, 
$\epsilon_i= 2.1$ is the lowest value of the dielectric constant. 

 Having the dielectric function of the three layers allows us to calculate  the  parallel component of the dielectric function to the interfaces  using effective medium theory as
\begin{equation}
 \epsilon_{||}(0)=\frac{2h}{L}\epsilon_i+\frac{L-2h}{L}\epsilon_{bulk}
 \label{effectivepar}
\end{equation}
Both equations (\ref{effectivebot},\ref{effectivepar}) are  the Maxwell-Wagner effective medium models for polar dielectric layered media. 

Based on the data reported by Fumagalli et al.\cite{Fumagalli}, we plot in Fig. (2) the best fit to the values of $\epsilon_{\bot}(0)$ as a function of the confinement $L$ given by Eq.(\ref{effectivebot}). Also, we plot the predicted parallel component of the static dielectric function $\epsilon_{||}(0)$. 
\begin{figure}[h]
\centering
  \includegraphics[height=6cm]{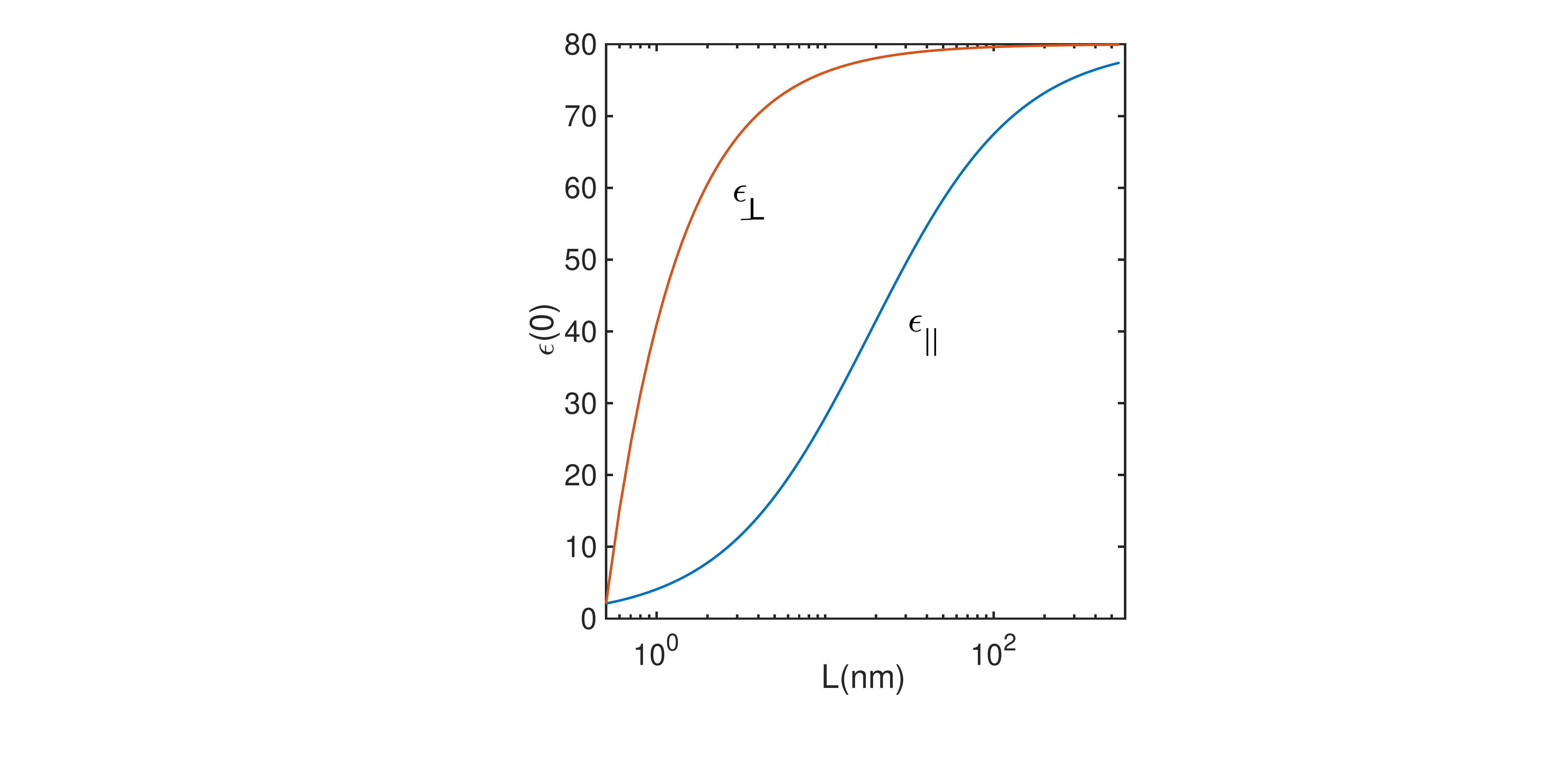}
  \caption{Perpendicular $\epsilon_{\bot}$ and parallel $\epsilon_{||}$ static dielectric functions, for different confining separations. The curves correspond to Eq.(\ref{effectivebot}) and Eq. (\ref{effectivepar}).}
  \label{static}
\end{figure}

Confinement only affects the static value of the dielectric function. This will be the anzats of this work.   For each confining length $L$, the  zero frequency contribution to the dielectric function comes from the experimental data  and for Matsubara frequencies $n\geq1$ the values of the dielectric function are evaluated using Eq. (\ref{water}) . This last term is independent of $L$.
In Figure (3) the perpendicular dielectric function is shown as a function of the Matsubara frequency number $n$ for different confining length $L$.  As explained, the $n=0$ changes for different confinement lengths, and the orange line is the non-static response, that does not changes with $L$.

\begin{figure}[h]
\centering
  \includegraphics[height=6cm]{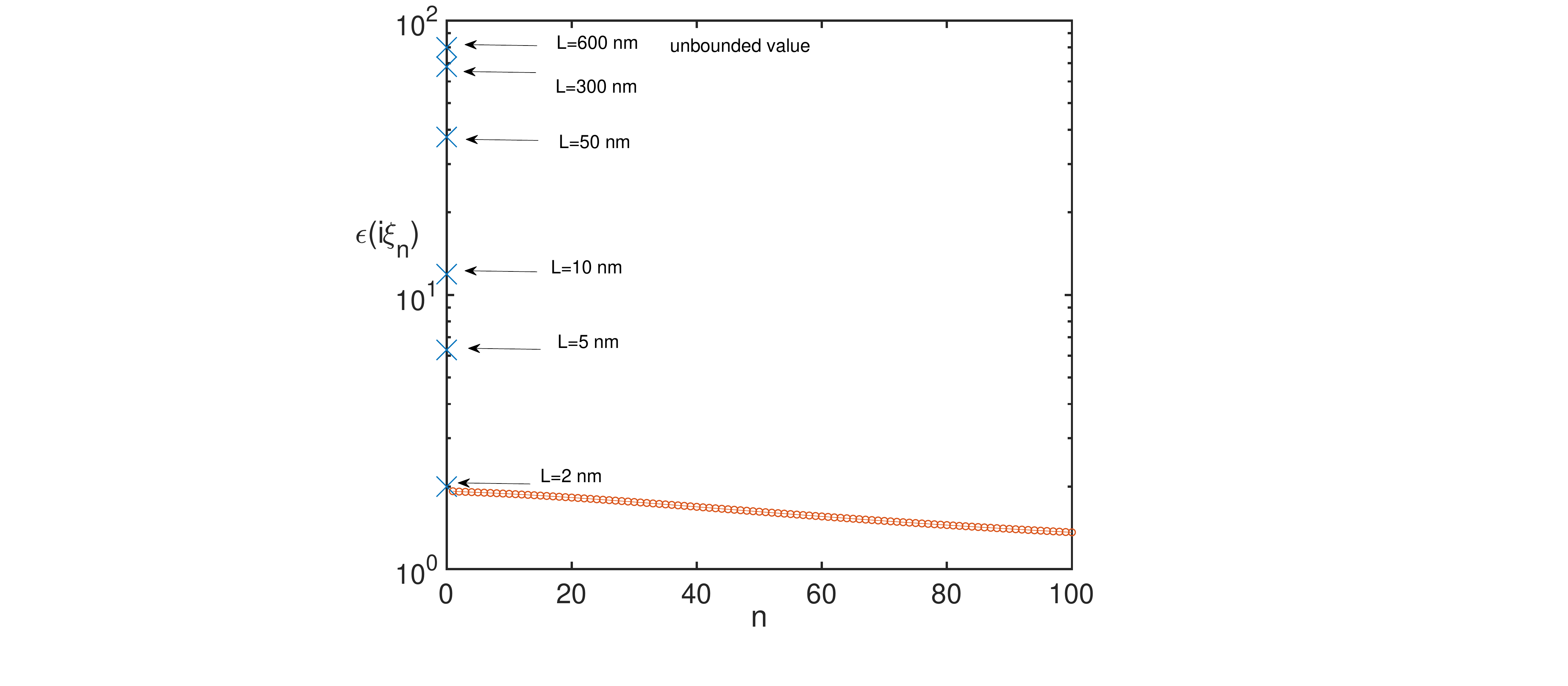}
  \caption{Dielectric function for the perpendicular dielectric function for different values of $n$, as a function of the confinement $L$. Only the $n=0$ changes while the contribution for $ngeq1$ is the same for all values of $L$ (orange line).
  }
  \label{waterf}
\end{figure}

The van der Waals energy between the two surfaces is
 \begin{equation}
    U=-\frac{A_H}{12\pi L^2},
    \label{energy}
\end{equation}
where $A_H$ is the Hamaker constant that is calculated using Lifshitz theory for generalized van der Waals forces.   In the non-retarded limit, the Hamaker constant $A_H$ between the surfaces is
\begin{equation}
A_H=-\frac{3KT}{2}\sum \limits_{n=0}^{\infty}\;^{'}\int_0^{\infty} k  ln(1-\Delta(n)^2 e^{-2kL}) dk,
\label{hamaker1}
\end{equation}
 where the prime in the sum over $n$ in Eq.(\ref{Ham}) means that the term $n=0$ has to be multiplied by $1/2$.  In Eq.(6), the normal component of the wave vector is $k$ and $L$ the separation between the walls.   The term $\Delta(n)$ is the normal incidence reflection coefficient  
\begin{equation}
   \Delta(n)=\frac{\epsilon_s(\zeta_n)-\epsilon(\zeta_n)}{\epsilon_s(\zeta_n)+\epsilon(\zeta_n)}, 
   \label{delta}
\end{equation}
and $\epsilon(\zeta_n)=\sqrt{\epsilon_{||}(\zeta_n)\epsilon_{\bot}(\zeta_n)}$.

For small separations, the integral in Eq. (\ref{hamaker1}) is calculated by expanding the log function in a Taylor series to obtain:
\begin{equation}
A_H=\frac{3KT}{2} \sum \limits_{n=0}^{\infty}\;^{'}\sum_{j=1}^{\infty}\frac{\Delta(n)^{2j}}{j^3}.
\label{Ham}
\end{equation}
To emphasize the role of the static value of the dielectric function or the $n=0$ term, we rewrite Eq.(\ref{Ham}) in terms of the polylogarithmic function $Li_n(x)$
whose argument satisfies $|x|<1$. Thus, Eq.(\ref{Ham}) is 

\begin{equation}
A_H=\frac{3KT}{2}\left(\frac{\epsilon_{||}(0)}{\epsilon_{\bot}(0)}\frac{Li_0(\Delta(0))}{2}+\sum_{n>0}^{\infty}Li_n(\Delta(n))\right).
\label{ham2}
\end{equation}

To evaluate the effect of the confinement in the Hamaker constant, first we consider that the confining plates are made of SiC. In Fig. (4), we plot the ratio $A_H/A_{Hb}$ as a function of the confining length $L$, where $A_{Hb}$ is the Hamaker constant of bulk or unbounded water.
The anomalous reduction of the dielectric function implies also a reduction in the Hamaker constant. In the inset of this figure, we plot again $A_H/A_{Hb}$ but when the confining walls are made of Au. In this case, there is a negligible effect due to confinement, since at zero frequency, the dielectric function of metals goes to $-\infty$ and the values of $\Delta(0)=1$. Thus, at  the nanoscale, in processes where van der Waals forces are important, such as self-assembly of Au nanoparticles \cite{Young2240}, changes of the static dielectric function are not important.

\begin{figure}[h]
\centering
  \includegraphics[height=6cm]{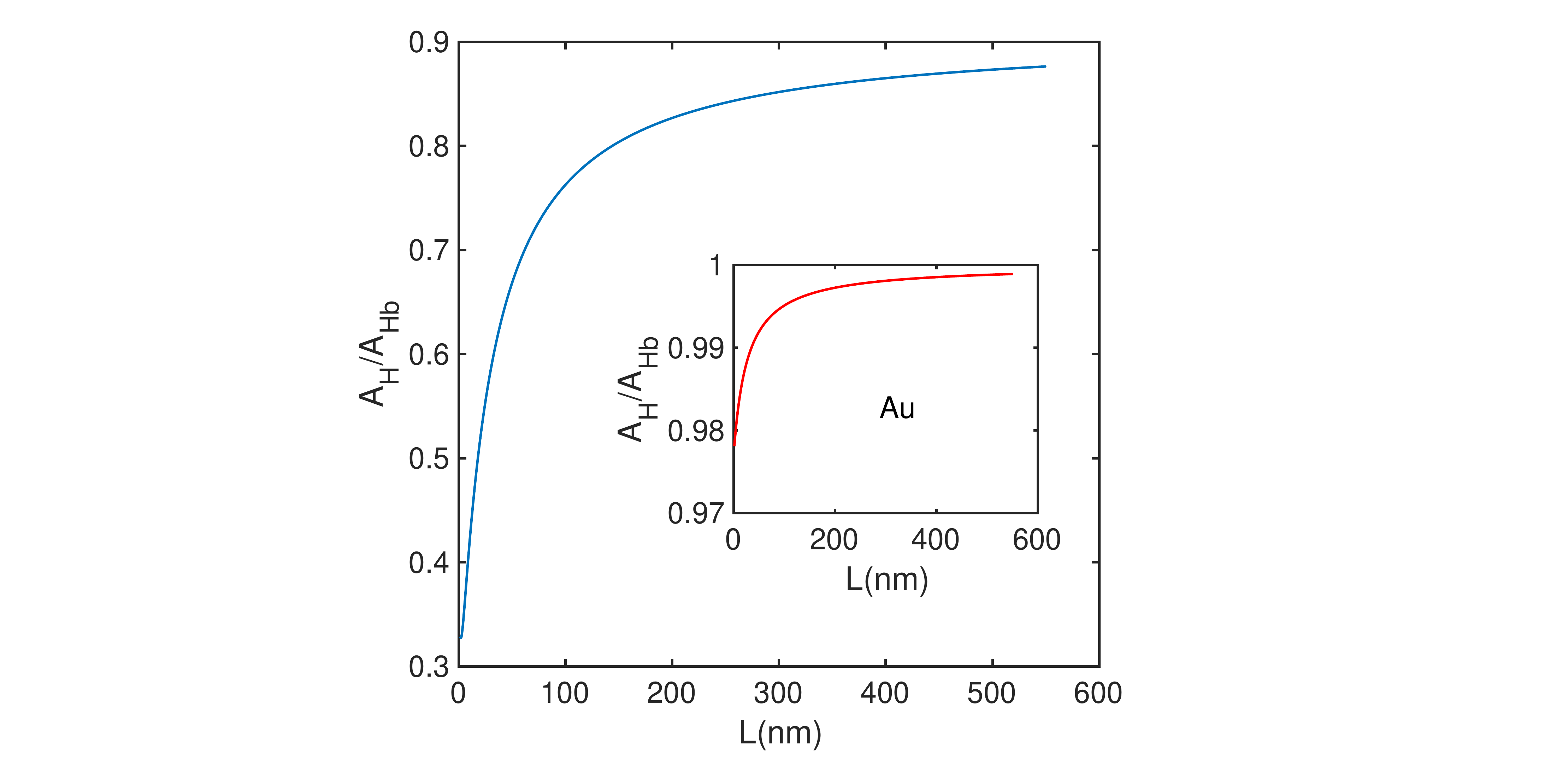}
  \caption{Ratio of the bounded Hamaker constant and the bulk constant $A_H/A_{Hb}$ as a function of the confinement length $L$. The confining walls are SiC. The inset is the same ratio when the confining walls are Au, and confinement has little effect in the Hamaker constant. }
  \label{hamaker}
\end{figure}

Confinement changes the dielectric function of water and also of electrolytes.  The role of electrolytes in Van der Waals interactions is also important since they have a strong screening effect, in particular for the zero-Matsubara frequency term \cite{Ningham72,Misra,C0SM00040J,Quesada,Jancovici,Petrache7982}.  
In an electrolyte solution the zero-term of  Hamaker constant in Eq.(\ref{hamaker1}) is \cite{Ningham72}
\begin{equation}
A_{H0}=-\frac{3KT}{4}\int_0^{\infty}k  ln(1-\Delta(0)^2 e^{-2sL}) dk,
\label{hamaker2}
\end{equation}
where the exponent now depends of the modified wave number $s=\sqrt{k^2+k_D^2}$ in the electrolyte. The Debye-H{u}ckel wavevector is defined in terms of the Deby length $k_D=1/\lambda_D$, 
\begin{equation}
\lambda_D=\sqrt{\epsilon}\sqrt{\epsilon_0\frac{KT }{\sum_i (q z_i)^2 c_i}},
\label{debye}
\end{equation}
with the usual definitions of the ion valency $z_i$, number concentration $c_i$, $q$ the value of the charge and $\epsilon_0$ the permittivity of vacuum.  The dielectric function of the solvent is $\epsilon$, water in our case, and the index $i$ is to label the different ions present.
 In the presence of  electrolytes, the factors $\Delta(0)$ in Eq. (\ref{delta}) are now written in terms of the modify wave number $s$ as \cite{Ningham72,NINGHAM97,parsegian_2005}
\begin{equation}
   \Delta(0)=\frac{s\epsilon_s(\zeta_0)-k\epsilon(\zeta_0)}{s\epsilon_s(\zeta_0)+k\epsilon(\zeta_0)}.
   \label{delta0}
\end{equation}
Thus, the Hamaker constant will be calculated using Eq. (8), but replacing the $n=0$ term with Eq.(\ref{hamaker2}).  

To study the combined effect of confinement and electrolytes we calculate the Debye length at $T=298$K for NaCl. In Figure 5(a) we show a contour plot of the Debye length as a function of the confinement length $L$ (horizontal axis), and of the concentration $c$ (M). Each line corresponds to different values of $\lambda_D$, with some values indicated in the labels. In Figure 5(b) we fix the concentration to $c=0.6(M)$ and plot the Debye length as a function of confinement separation $L$ (nm). We observe that as $L$ decreases, the Debye length decreases too. Thus, even at a fixed concentration of electrolytes the confinement will induce changes in $\lambda_D$. Only low concentrations are considered to avoid other effects at higher salt concentrations as reported by Smith \cite{Smith}. 

\begin{figure}[h]
\centering
  \includegraphics[height=6cm]{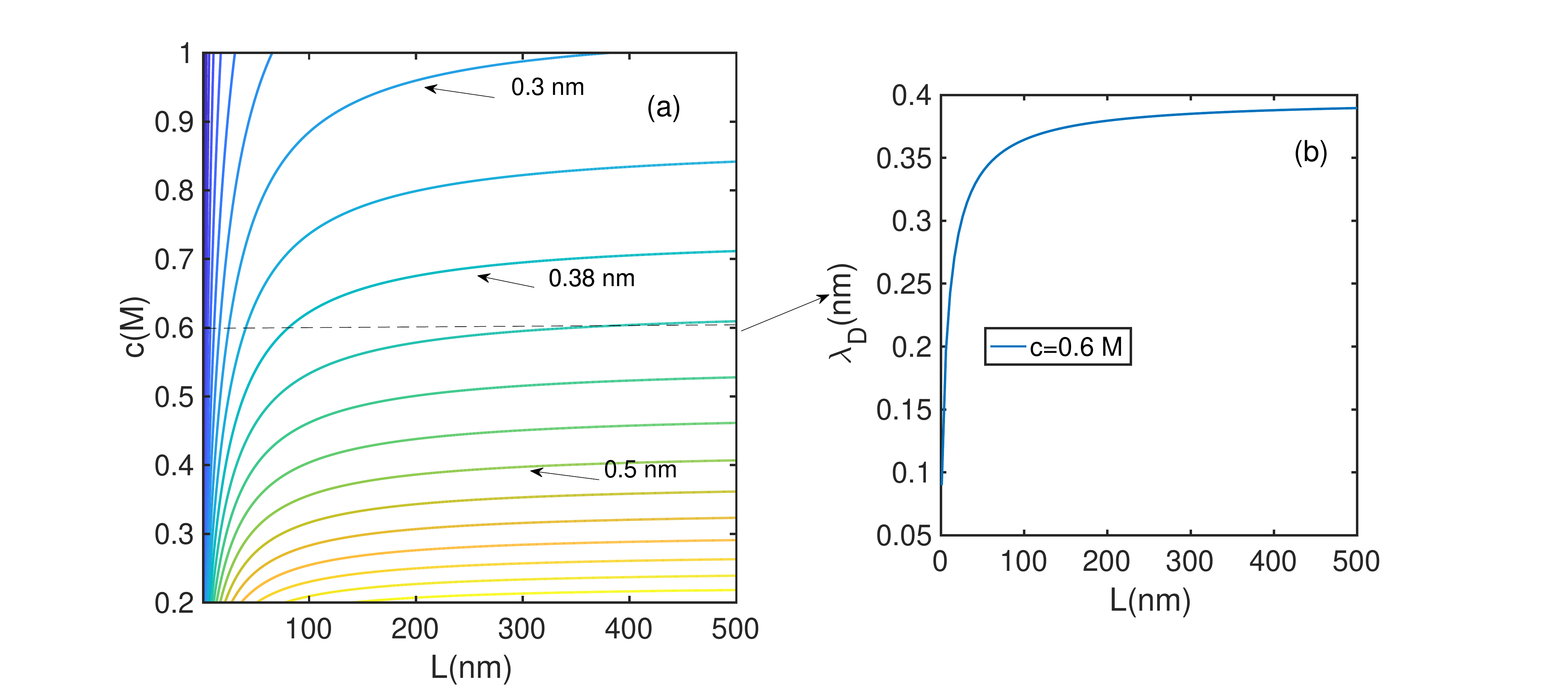}
  \caption{(a) Contour plot of the Debye length $\lambda_D$ for different molar concentrations of NaCl as a function of the confinement distance $L(nm)$. The labels on some of the iso-curves are shown. Each curve is a different value of the Debye length. (b) For a fixed concentration of $c=0.6$ (M) we plot the Debye length as a function of confinement.  }
  \label{hamaker}
\end{figure}

The changes of the Hamaker constant with electrolytes, as a function of the confinement length, is shown in Fig. (6). As before we compare with the Hamaker constant of bulk or unconfined water $A_{Hb}$. 

\begin{figure}[h]
\centering
  \includegraphics[height=6cm]{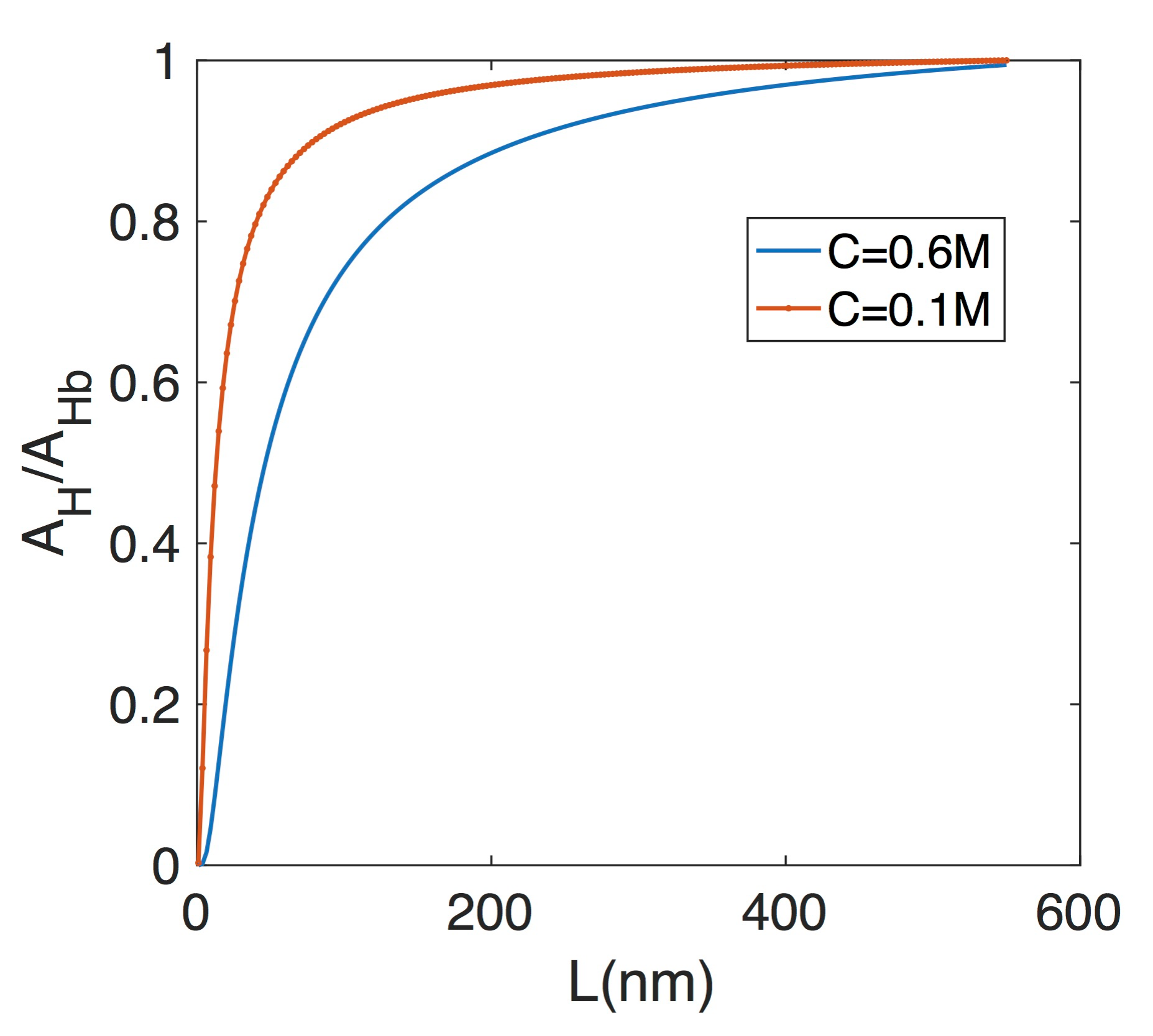}
  \caption{Changes in the Hamaker constant $A_{H}/A_{Hb}$ as a function of the confinement length $L$ for two different electrolyte concentrations. As before $A_{Hb}$ is the constant taking the bulk value of the dielectric function of water. As the confinement length decreases, the Hamaker constant decreases while the concentration is kept fixed. }
  \label{hamaker}
\end{figure}

\section{Conclusions}

 The anomalous reduction of the static dielectric function of water has a strong effect on the Van der Waals forces. If the confining walls are made of a dielectric material, the Hamaker constant decreases as the confining length decreases. For metallic confining walls, the effect of confinement has little effect, since the static value of the dielectric constant of metals is large.  Also, we show that the Deby length can be reduced by keeping constant the salt concentration and changing the confinement length.  Since the value of the Hamaker constant also depends on the electrolytic concentration,  changing the confinement length again has an important effect on  the Hamaker constant. Confinement decreases the Hamaker constant  significantly, having potential applications in nanofluidic devices where the van der Waals interaction can lead to unwanted pull-in or jump to contact effects \cite{Das_2009,Esq-mems}, in particular where electrolytes are involved \cite{BOYD2011387}.
 \begin{acknowledgement}
The author thanks  Guadalupe Goméz Farfán,  Giuseppe Pirruccio and Juan V. Escobar for helpful discussions and help with the art work. 
\end{acknowledgement}

\providecommand{\latin}[1]{#1}
\makeatletter
\providecommand{\doi}
  {\begingroup\let\do\@makeother\dospecials
  \catcode`\{=1 \catcode`\}=2 \doi@aux}
\providecommand{\doi@aux}[1]{\endgroup\texttt{#1}}
\makeatother
\providecommand*\mcitethebibliography{\thebibliography}
\csname @ifundefined\endcsname{endmcitethebibliography}
  {\let\endmcitethebibliography\endthebibliography}{}

\end{document}